\documentstyle[11pt,epsfig]{article}
\begin{document}
LYCEN 9623 July 96

nucl-th/9607051

\begin{center}
{\bf PIONS IN THE NUCLEAR MEDIUM}
\footnote{\it Talk given at the Workshop MESON'96, 
Cracow, Poland, 10-14 may 96}
 
\vskip 1 true cm
 
G. CHANFRAY
\vskip 0.5 true cm
{\it IPN-Lyon, 43 Bd. du 11 Novembre 1918,
F-69622 Villeurbanne C\'{e}dex, France.}
\end{center}
\vskip 1true cm

\begin{center}
{\bf Abstract}
\end{center}
{\it We discuss various aspects of pion physics in the nuclear medium. 
We first study s-wave pion-nucleus interaction in connection with 
chiral symmetry restoration and
quark condensate in the nuclear medium. We then adress the question of p-wave 
pion-nucleus interaction and collective pionic modes in nuclei 
and draw the consequences for in medium $\pi\pi$ correlations 
especially in the scalar-isoscalar channel. We finally discuss the modification
of the rho meson mass spectrum at finite density and/or temperature in connection 
with relativistic heavy ion collisions.  }
%%%%%%%%%%%%%%%%%%%%%%%%%%%%%%%%%%%%%%%%%%%%%%%%%%%%%%%%%%%%%%%%%%%
\section{Introduction}
%%%%%%%%%%%%%%%%%%%%%%%%%%%%%%%%%%%%%%%%%%%%%%%%%%%%%%%%%%%%%%%%%%%
The physics of  pions in the nuclear medium has already a long story. 
This domain contains two complementary aspects. The first one concerns 
the use of 
pions as a probe of the nucleus and the second one is devoted to the key role 
of the pion in the nuclear many-body problem. Most of the 
considerable knowledge  which has been accumulated can be found in \cite{BIBLE}.
In this paper we will show that all these results are of prime importance 
for the understanding of new problematics of present day hadronuclear physics. 
In that respect we will discuss in section $2$ the connection between chiral
symmetry restoration and the s-wave pion-nucleus interaction. 
 Section $3$ is devoted to the physics of collective pion-delta modes, 
 sometimes called pionic branch or pisobars. We discuss the consequences 
 of these collective phenomena on two-pion propagation and correlations 
 in nuclear matter (section $4$). Finally we examine 
 in section $5$ the role of interacting
 $\pi N \Delta$ configurations in highly excited matter 
 produced in relativistic 
 heavy ion collisions; special emphasis will be put on the rho meson 
 whose mass spectrum may reveal significative modification 
 of the vacuum structure at finite  density and/or temperature

%%%%%%%%%%%%%%%%%%%%%%%%%%%%%%%%%%%%%%%%%%%%%%%%%%%%%%%%%%%%% 
\section{ Chiral symmetry restoration and s-wave pion-nucleus
 interaction}
%%%%%%%%%%%%%%%%%%%%%%%%%%%%%%%%%%%%%%%%%%%%%%%%%%%%%%%%%%%%%%%%%%%%%%%%%%%

Asymptotic freedom and color confinement are usually considered
as the most prominent properties of our theory of strong interaction,
Quantum Chromodynamics (QCD). However
QCD also  possesses an almost exact symmetry, the $SU(2)_R\otimes SU(2)_L$
chiral symmetry which is certainly the most important key for the
understanding of many phenomena in low energy hadron physics.
This symmetry originates from the fact
that the QCD Lagrangian is almost invariant under the separate flavor $SU(2)$
transformations  of
right handed $q_R=(u_R,d_R)$ and left handed $q_L=(u_L,d_L)$
light quark fields $u$ and $d$
 
\begin{equation}
q_R\to e^{i\vec\tau.\vec\alpha_R/2}\,q_R\qquad
q_L\to e^{i\vec\tau.\vec\alpha_L/2}\,q_L
\end{equation}
The small explicit violation of chiral symmetry
is given by the mass term
of the QCD Lagrangian which is, neglecting isospin violation,
 
\begin{equation}
{\cal L}_{\chi SB}=-m_q\, (\bar u u+\bar d d)
\end{equation}
where the averaged light quark mass $m_q=(m_u+m_d)/2\le 10 MeV$,
the scale of explicit chiral symmetry breaking, has to be compared with typical
hadron masses of order $1\, GeV$, indicating that the symmetry is excellent.
The normal Wigner realization would imply a doubling of the flavor $SU(2)$
symmetry built on left-handed and right-handed quarks.
Each hadron should have a chiral partner with opposite parity and about the same
mass. This is obviously not the case since, for instance, the first
hypothetical chiral partner of the nucleon is the $S_{11}$ resonance
whose mass is larger than $1.5\, GeV$. However a symmetry present at the level
 of
the Lagrangian is not necessarily realized in the spectrum. This is precisely
 what
happens for chiral symmetry which is realized in the Goldstone manner or,
equivalently stated, is spontaneously broken. One order parameter of this
 breaking in
the QCD vacuum is provided by the quark condensate
 
\begin{equation}
<\bar q q>_{vac}={1\over 2}<\bar u_R u_L+
\bar u_L u_R+\bar d_R d_L+ \bar d_L d_R>\simeq -(220\, MeV)^3
\end{equation}
 which is  an observable which mixes left-handed and right-handed quarks.
The associated Goldstone boson is the pion whose small mass comes from the small
explicit chiral symmetry breaking. There is a very important relation relating
the explicit and spontaneous breaking at the quark scale and at the hadronic
scale namely the Gell-Mann-Oakes-Renner  relation \cite{GOR} valid to
lowest order in the explicit symmetry
breaking parameters ($m_q, m^2_\pi)$
 
\begin{equation}
-2 m_q <\bar q q>_{vac}=m^2_\pi f^2_\pi
\end{equation}
where the pion decay constant $f_\pi$ plays the role of the order parameter for
 the
spontaneous breaking at the hadronic level. The dynamical origin of this
 spontaneous
breaking, highly non pertubative in nature, is not yet fully understood. However,
we have very good reasons to believe (based on lattice simulation or
effective hadron theories) that the quark condensate decreases, indicating
partial restoration of the symmetry, with increasing baryonic density
and/or temperature. Hence, before the full restoration at some critical
density $\rho_c$ and critical temperature $T_c\simeq 150-200 MeV$ we expect
that the mesons, which are the first excitations of the QCD vacuum,
will be appreciably modified.

Already in nuclear matter at normal density $\rho_0$ there is a
sizeable restoration of chiral symmetry whose origin can be understood in
a very simple picture. A nucleon embedded in nuclear matter can be seen as
a bubble in the QCD vacuum in which chiral symmetry is restored. Therefore
on the average, the quark condensate will decrease. This qualitative statement
can be made quantitative with the first order result \cite{COH},
derived with help of the GOR relation
 
\begin{equation}
R_{first\, order}={<\bar q q(\rho)>\over <\bar q q>_{vac}}=
1-{\rho \Sigma_N \over f^2_\pi m^2_\pi}
\label{first}
\end{equation}
Hence, to leading order in density the amount of chiral symmetry
restoration, given by the ratio $R$, is  governed by the pion-nucleon sigma
commutator $\Sigma_N=<N|\big[Q^5_i,[Q^5_i,H]\big]|N>\simeq 45 MeV$,
which is a measure of the explicit symmetry breaking on the nucleon.
Putting the numbers together we find a $30 \%$ restoration at normal
nuclear matter density or in the interior of heavy nuclei. This result, valid
 for a
non-interacting medium, can be promoted to an exact result \cite{GC1} provided
the $\pi N$ sigma commutator is replaced by the full $\pi$-Nucleus
sigma commutator
(per nucleon) $\tilde \Sigma_N(\rho)$
 
\begin{equation}
R={<\bar q q(\rho)>\over <\bar q q>_{vac}}=
1-{\rho \tilde \Sigma_N (\rho)\over f^2_\pi m^2_\pi}
\label{exact}
\end{equation}
 
\begin{equation}
\tilde\Sigma_N(\rho)={1\over A}\,<A|\big[Q^5_i,[Q^5_i,H]\big]|A>
\end{equation}
An alternative form of the exact result (\ref{exact}) can be obtained by
introducing the interpolating PCAC pion field related to the divergence of the
axial current according to $\varphi_i=-\partial_\mu {\cal A}^\mu_i/f_\pi
 m^2_\pi$.
There is a venerable low energy theorem \cite{VEN} which states that the double
sigma commutator on any target is proportionnal to the total isoscalar
 scattering
amplitude between the target and PCAC soft pion {\it i.e.} a pion with a
 vanishing
energy momentum. As a consequence, it can be shown that
 equ.(\ref{exact}) takes the equivalent form \cite{GC2}

\begin{equation}
R={m^2_\pi\over m^2_\pi + \Pi(0)}=1-{\Pi(0)
\over m^2_\pi \,+\,\Pi(0)}
\label{PCAC}
\end{equation}
where $\Pi(0)$ is the full PCAC pion self-energy at four momentum $q=0$
and corresponds physically to the soft pion-nucleus optical potential. The advantage
of this formulation, which is at least to me more intuitive,
 is that all the mechanisms governing chiral symmetry
restoration can be described in terms of contributions to the full soft
 pion-nucleus
amplitude per unit volume $T=\Pi(0)/[ 1 +\Pi(0)/m^2_\pi]$. Keeping only in $T$
the Born term $\Pi(0)$
limited to first order in density, namely $T^0_{Born}=\Pi^0(0)=\rho \Sigma_N /
 f^2_\pi$, we recover back the first order result (\ref{first}).
 However, the full T matrix deviates from the optical potential $\Pi(0)$
 by coherent rescattering
 (i.e the denominator in T). The fact that the first order optical potential
 $\Pi^0$
 is repulsive may lead to the conclusion that coherent rescattering hinders
 chiral symmetry restoration. In other words, as proposed by
 M. Ericson \cite{ME1}, there is a possible reaction of the
 medium against chiral symmetry restoration.

 In two recent papers \cite{GC1,DEL} we have studied how much this conclusion
 might be altered
by higher order contributions to the irreducible pion self energy $\Pi(0)$ such
 as
pion exchange effects or the influence of short range correlations   combined
with higher order exchange effects.
 
The effect of pion exchange  can be obtained quantitatively
by the direct evaluation
of the ground state expectation value
$<H_{\chi SB}>= (1/2)<A|\int d{\bf r}\, m^2_\pi \vec \varphi . \vec\varphi
({\bf r})|A>$ on a correlated ground state wave function or, in the PCAC
 picture,
by calculating the scattering amplitudes, such as the one depicted in fig 1a,
within the standard  non linear chiral Lagrangian. We
found \cite{GC1} a slight acceleration of chiral
symmetry restoration given by $\delta R\simeq -0.04\, (\rho / \rho_0)^2$,
 a $4\%$ effect at  normal nuclear matter density, closely
linked to the pion excess.
 
The incoherent rescattering represents the rescattering of the soft pion in
presence of short range correlations represented
by a double wavy line on fig.1b. The
intermediate pion of fig.1b acquires a momentum $q_c\simeq 1/R_c\simeq m_\omega
\simeq 700 MeV$ where $R_c$ is the size of the pair correlation hole.
Consequently, the corresponding contribution to the pion self energy possesses
a piece going as $q_c^2$
in manifest conflict with chiral symmetry.
We have shown \cite{DEL}, using a linear sigma model (fig.1c) or non linear
sigma model (fig.1d), that higher order corrections exactly cancel
this $q_c^2$ piece, as it should be.
 The net effect of short range correlations is thus weak.
Their effect becomes  negligibely small if the  scalar-isoscalar radius $R_S$,
much larger than the correlation hole radius $R_c$, is taken into account.
 
In the linear sigma model, the remaining effect of higher order contributions
 such
as depicted on fig.1c
is extremely large. For instance Birse and Mac Govern \cite{BIRSE}
found a contribution

\begin{equation}
\delta \Pi= {5\over 2} m^2_\pi\left({\rho\Sigma_N\over f^2_\pi m^2_\pi}\right)^2
\qquad\to\qquad
\delta R= -{5\over 2}\left({\rho\Sigma_N\over f^2_\pi m^2_\pi}\right)^2\simeq-
0.25 \left({\rho\over \rho_0}\right)^2
\end{equation}
but such a result is ruled out by standard $\pi$-nucleus phenomenology
yielding for instance a unrealistically large $(b_0)_{eff}$. To look,
in a more realistic way, at the
inluence  of higher order terms we start from the non linear Lagrangian
 
\begin{equation}
{\cal L} =i\bar \psi \gamma^\mu \partial_\mu \psi- g\bar\psi
(\sigma + i\vec \tau .\vec\pi
\gamma_5)\psi\,+\, {1\over 2} \big(\partial_\mu \sigma \partial^\mu \sigma
+\partial_\mu \vec \pi . \partial^\mu \vec \pi \big)
+f^2_\pi m^2_\pi\, \sigma -{\Sigma_N\over f^2_\pi}\,\sigma\,
\bar \psi(\sigma + i\vec \tau .\vec\pi \gamma_5)\psi
\label{real}
\end{equation}
with $\sigma=(f^2_\pi-\vec \pi^2)^{1/2}$. We have explicitely included
in eq.(\ref{real}) a chiral
symmetry breaking piece proportionnal to $\Sigma_N$ to obtain the correct
pion-nucleon sigma term at the tree level while maintaining the QCD algebraic
 identity
$H_{\chi SB}=\big[Q^5_i,[Q^5_i,H]\big]$.
For convenience we introduce a new nucleon field and the PCAC pion field
according to
 
\begin{equation}
N=\left({\sigma+i\vec\tau .\vec\pi \gamma_5\over f_\pi}\right)^{1/2}\,\psi
\qquad
\vec\varphi=\vec\pi \left(1-{\Sigma_N\, \bar N N\over f^2_\pi m^2_\pi}\right)
\end{equation}
This generates a new form for the chiral symmetry breaking piece \cite{DEL}
 
\begin{equation}
{\cal L}_{\chi SB}=f^2_\pi m^2_\pi
\left(1-{\Sigma_N\, \bar N N\over f^2_\pi m^2_\pi}\right)
\,-\,{1\over 2}m^2_\pi{\varphi^2\over
1-\Sigma_N\, \bar N N / f^2_\pi m^2_\pi}\,+\, O(\varphi^4)
\end{equation}
From the above equation we can directly read off the result for the
soft pion optical potential
 
\begin{equation}
T^{Born}_{soft}=\Pi(0)={\rho \Sigma_N /f^2_\pi\over 1-\rho \Sigma_N /f^2_\pi
 m^2_\pi}
\end{equation}
We see that the first order optical potential is renormalized by a set of
higher order contact terms such as displayed on fig.1d.
The distorsion of the soft pion wave through coherent rescattering modifies
the Born amplitude in such a way that the free nucleon amplitude is recovered
 
\begin{equation}
T={\Pi(0)\over 1 +\Pi(0)/m^2_\pi}={\rho\Sigma_N \over f^2_\pi}
\end{equation}
One can also notice that this result could have been directly obtained
 by taking the expectation
value of $-{\cal L}_{\chi SB}$. Hence coherent rescattering is exactly cancelled
 by
the contact terms, contrary to what happens in the linear sigma model
 where it is overcompensated.
 
From the above discussion one can draw the following conclusions~:
 
- The role of short range correlations in chiral symmetry restoration is
always very small.
 
- The reaction of the medium against chiral symmetry restoration, associated
to coherent rescattering, seems to be always compensated by higher order
exchange contributions
 
- The main deviation of the nuclear sigma term with respect to the bare nucleon
sigma term originates from pion exchange associated to pion excess in nuclei.
Nevertheless, 
some higher order terms in the density may very well be present in the effective
chiral Lagrangian governing low energy hadron and nuclear physics. These terms
still largely unknown should be determined by experiments and may alter this
 last conclusion. However, an experimental information on the nuclear sigma 
 term can be obtained through a Fubini-Furlan type analysis whose only input are 
 scattering data. As shown in \cite{GC1} the deviation of the nuclear sigma term 
 with respect to the bare pion-nucleon sigma term lies in the range $3-9\, MeV$ 
 (depending on the assumptions on a dispersive correction) and is compatible 
 with the calculated pion exchange contribution.
 
\begin{figure}
  \begin{center}
    \mbox{\epsfxsize=9cm
          \epsfbox{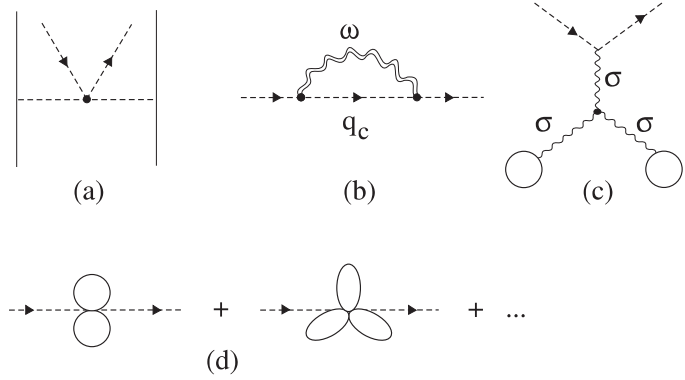}}
  \end{center}
  \caption{Examples of medium correction to the
pion-nucleus sigma term governing chiral symmetry
restoration.The external dashed lines represent soft pions.
 (1a) pion exchange. (1b) Incoherent rescattering in presence of short
range correlations.
(1c) Higher order exchange in the linear sigma model.
(1d) Higher order contact terms in the non linear
sigma model.}
  \label{figure1}
\end{figure}

%%%%%%%%%%%%%%%%%%%%%%%%%%%%%%%%%%%%%%%%%%%%%%%%%%%%%%%%%%%%%%%%%%%%%%%%%%%%%%%%%%%%%%%
\section {Collective pionic modes and p-wave pion-nucleus interaction}
%%%%%%%%%%%%%%%%%%%%%%%%%%%%%%%%%%%%%%%%%%%%%%%%%%%%%%%%%%%%%%%%%%%%%%%%%%%%%%%%%%%%%
The propagation of a high energy pion (say $\omega > m_\pi$) is modified mainly 
by its p-wave self-interaction in the nuclear medium. The pion propagator has 
the form~:
\begin{equation}
D_\pi({\bf k},\omega)\,=\,\big[\omega^2-\omega^2_k- S({\bf k},\omega)]^{-1}  
\end{equation}
The p-wave self-energy is dominated by the virtual $\Delta h$ excitations and 
it is  extremely important to incorporate the effect of the repulsive 
short-range correlations through the introduction of the $g'$ parameter. 
In a more refined 
description (see below) the coupling of the pion to $p h$ and $2p 2h$ excitations 
should also be included.  It is possible to understand the main aspects of pion 
propagation and pionic collective modes within a very simple two-levels 
model \cite{CASN}. Keeping only 
the coupling  to $\Delta h$ states in the static approximation ($\epsilon_{\Delta k}
=\sqrt{k^2+M^2_\Delta}-M_N$) and neglecting the delta width, 
the pion self-energy has the form~:
\begin{equation} 
S({\bf k},\omega)\,=\,k^2\,\tilde\Pi^0({\bf k},\omega)\,=
\,k^2\,\Pi^0({\bf k},\omega)\,/\,\left(1-g'_{\Delta\Delta}
\Pi^0({\bf k},\omega)\right)
\end{equation}
Here $g'_{\Delta\Delta} \simeq 0.5$ accounts for the short range screening of 
the polarisation bubble $\Pi^0$ which has the usual form~:
\begin{equation}
\Pi^0({\bf k},\omega)\,=\, 
{4\over 9}\left({f^*_{\pi N \Delta}\over m_\pi}
\Gamma(k)\right)^2\,\rho\,\left({1\over \omega-\epsilon_{\Delta k}+i \eta}
-{1\over \omega+\epsilon_{\Delta k}}\right)
\end{equation}
It is a simple matter to show that the full pion propagator takes the following form
\begin{equation}
D_\pi({\bf k},\omega; \rho)\,= {Z_1(k,\omega ;\rho)\over \omega^2 -
\Omega_1^2( k,\omega; \rho) + i \eta}\,
+\,{Z_2(k,\omega ;\rho)\over \omega^2-
\Omega_2^2(k,\omega; \rho) + i\eta}
\end{equation}
\begin{figure}
  \begin{center}
    \mbox{\epsfxsize=7cm
          \epsfbox{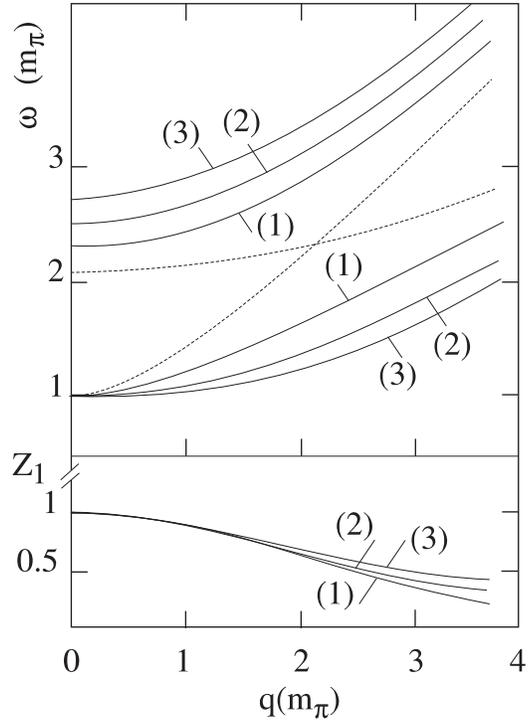}}
  \end{center}
  \caption{Dispersion curves for $\Omega_1$, $\Omega_2$ and $Z_1$ weight factor
    for various values of $\rho / \rho_0$. The free pion and delta branches 
    are also shown. }
   \label{figure2}
\end{figure}

Due to its coupling to  the $\Delta$ it is clear that the $\pi$ 
(and the $\Delta$ as well) is no longer an eigenstate of the system  but becomes a
mixture of two  $\pi \Delta$ modes with eigenenergy
$\Omega_1, \Omega_2$  ($\Omega_1 < \Omega_2$) and strength
$Z_1,Z_2$ ($Z_1+Z_2=1$) shown on figure 2. 
At low momentum the lower branch sometimes called the pionic branch is 
dominantly pionic whereas the upper branch is mainly made of $\Delta h$ states;
the two branches exchange their structure at momentum transfers of 
the order of $2-3 m_\pi$. The very important point for the following discussion 
(next section) is the smallness of the group velocity $d \Omega_1/ dk$
associated to the pionic branch at low momentum transfer and at 
sufficiently high density. In spite of its simplicity this model gives the position 
and the strength of these branches close to the results of 
much more detailed calculation \cite{DELG1}. 
\begin{figure}
  \begin{center}
    \mbox{\epsfxsize=9cm
          \epsfbox{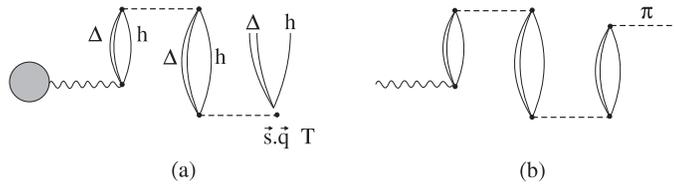}}
  \end{center}
  \caption{ (3a) Propagation of the longitudinal spin-isospin excitation 
  in the nuclear medium. (3b) Coherent pion.}
  \label{figure3}
\end{figure}

\begin{figure}
  \begin{center}
    \mbox{\epsfxsize=7cm
          \epsfbox{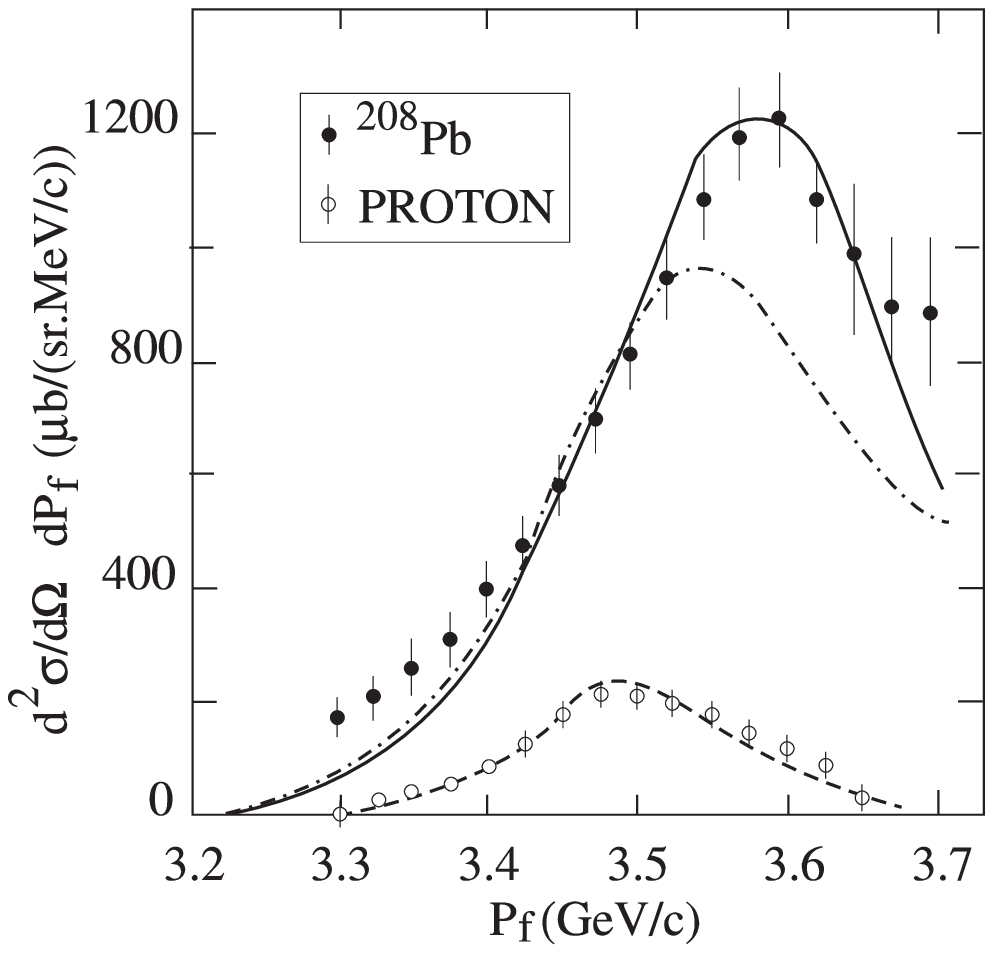}}
  \end{center}
  \caption{Comparison of theory to experiment for $\Delta$ excitation in 
  $^{208} Pb$ by the ($^3 He, T$) reaction at $2\, GeV$ and zero angle. 
  The dot-dashed curve is the first order response and the full curve includes
  collective $\pi\Delta$ effects. The dashed curve corresponds to a calculation 
  on a proton target.}
  \label{figure4}
\end{figure}

The lower (pionic) branch is mainly space-like and is in principle accessible by looking
at the longitudinal spin-isospin response with a well chosen probe. 
The propagation of the excitation in the medium through the pion  
builds up the collective response of the nucleus as shown in fig. 3a. 
Consequently,  
the delta-hole bump will be depopulated in favor of the pionic branch. 
Hence, a signature of these collective effects should be the downwards shift
of the delta peak with respect to its position in the free nucleonic response 
or in the transverse response measured with real and virtual photons. 
Despite their peripheral character, we believe that the ($^3 He, T$) experiment 
performed at SATURNE at $2\, GeV$ incident energy \cite{CONT} 
gives a clear indication in favor of the existence of these collective modes. 
The ($^3 He, T$) data on various nuclei ranging from carbon to lead show a 
systematic downwards shift of about $70\, MeV$ with respect to the proton data
(see fig. 4). Sophisticated calculations 
\cite{DELG2}  without free parameters 
(in the sense that all the needed ingredients have been extracted from  
phenomenology) have shown that the collective effects are absolutely needed
to explain the data (fig. 4); similar conclusions have been reached in 
\cite{UDA}.  The decay channels of these collective modes have also been studied
at SATURNE \cite{HEN}
and are still compatible with the above collective mechanism. The most direct evidence
of the real existence of these collective modes is probably 
the experimental observation of coherent pion production, the target nucleus 
remaining in its ground state, shown 
of fig. 3b. A new dedicated experiment is presently performed to study in detail 
coherent pion production; details on this last topic can be found in the papers of
B. Ramstein and L. Farhi in these proceedings.
  
%%%%%%%%%%%%%%%%%%%%%%%%%%%%%%%%%%%%%%%%%%%%%%%%%%%%%%%%%%%%%%%%%%%%%%%%%%%%%5    
\section{Pion-pion correlation and two-pion resonances in nuclear matter}
%%%%%%%%%%%%%%%%%%%%%%%%%%%%%%%%%%%%%%%%%%%%%%%%%%%%%%%%%%%%%%%%%%%%%%%%%%%%%%
Since the pion propagation properties are significantly modified in 
nuclear matter, one can also expect  two-pion systems such as two-pion 
resonances to be also appreciably modified. For this purpose let us consider 
the two-pion propagator in nuclear matter. For a two-pion system in its center 
of mass frame with relative momentum ${\bf k}$ and total energy $E$, it reads~:
\begin{equation}
G_{2\pi}({\bf k},E)=\int {idk^0\over 2\pi}\, D_\pi({\bf k},k_0)\, 
D_\pi(-{\bf k},E-k_0)
\end{equation} 
In the two level model, its explicit expression is ~:
\begin{equation}  
G_{2\pi}({\bf k},E)=\sum_{i,j=1}^2\,{\Omega_i(k)+\Omega_j(k) \over
2 \Omega_i(k)\, \Omega_j(k)}\, {1 \over E^2-\, \left(\Omega_i(k)+\
\Omega_j(k)\right)^2+ i \eta}
\end{equation}
Obviously in the zero density limit one recovers the bare two pion 
propagator namely $(E^2-4 \omega_k^2 + i\eta)^{-1}$. To construct 
the full effective interaction ( $T$ matrix) one starts from a certain model
for the $\pi \pi$ potential built with bare meson exchanges whose parameters are 
fitted on phase shifts and scattering lengths. The T matrix (invariant
scattering amplitude) is obtained as a solution of the Lippman-Schwinger 
equation
\begin{equation}
T({\bf q}, {\bf q}';\,E)\,=\,V({\bf q}, {\bf q}';\,E)+
\int {{\bf d k}\over (2\pi)^3}\,V({\bf q}, {\bf k};\,E)\,
G_{2\pi}({\bf k},E)\,T({\bf k}, {\bf q}';\,E)
\end{equation}
projected on the various channels such as $I=J=0$ ($\sigma$ channel) 
or $I=J=1$ ($\rho$ channel). A very important remark can be made~: at 
low energy the two pion strength distribution ({i.e.} $Im T$) receives 
a contribution of the type~:
\begin{equation}
\int\, {{\bf d k}\over (2\pi)^3}\, {Z_1(k)\over \Omega_1(k)}\,
\delta(E^2-4\,\Omega_1^2(k))
\end{equation}
which gives after phase space integration a factor 
$\big[d \Omega_1(k)/dk\big]^{-1}$. Since the group velocity $d \Omega_1(k)/dk$
may become very small at low momentum with increasing density, 
an accumulation of strength near the $E=2 m_\pi$ threshold can be produced. 
This very simple argument 
can be strictly applied only in the sigma channel since gauge corrections
introduce other medium effects in the rho channel; this last point 
will be discussed in the next section.

The in medium pion-pion $T$ matrix has been studied first in two 
phenomenological models. In the simplest one, described in \cite{CASN} the 
$\pi \pi$ interaction is described by a simple scalar exchange 
roughly simulating the coupling to the $K \bar K$ channel. 
The other model is the much more sophisticated J\"{u}lich model \cite{LOHSE} 
which explicitely incorporates the coupling to the $K \bar K$ channel. 
Although very different, these two models give very similar results for
the $I=J=0$ channel.
Both predict a very sharp structure near two-pion threshold with even 
the occurence of a two-pion $I=J=0$ 
bound state for a density slightly below 
normal nuclear matter density. It is important to check wether this effect survives
when the width of the in medium quasi-pion is incorporated. This can be done
by making the substitution
\begin{equation}
\Omega_j(k)\,\to\,\Omega_j(k)\,+\,i \,{Im \tilde \Pi^0 \big({\bf k},
\Omega_j(k)\big)\over 2\, \Omega_j(k)} 
\end{equation}
where $Im \tilde \Pi^0 $, calculated along the line $j$, takes into account
the delta width, corrected from Pauli blocking \cite{OSET}, together 
with extra $2p 2h$ contributions not reducible to a delta width piece
\cite{AOUSS}. The imaginary part of the two-pion propagator is obtained 
through a spectral representation~:
\begin{equation}
Im \, G_{2\pi}=-{1\over \pi}\,\sum_0^E\,d\omega\, Im D_\pi ({\bf k},E)\,
Im D_\pi({\bf k}, E-\omega) 
\end{equation}
and the real part is calculated using dispersion relation. The results, 
with widths included,  is depicted on fig. 5 for the first model. 
It appears that the sructure near two-pion treshold is still there. 
\begin{figure}
  \begin{center}
    \mbox{\epsfxsize=11cm
          \epsfbox{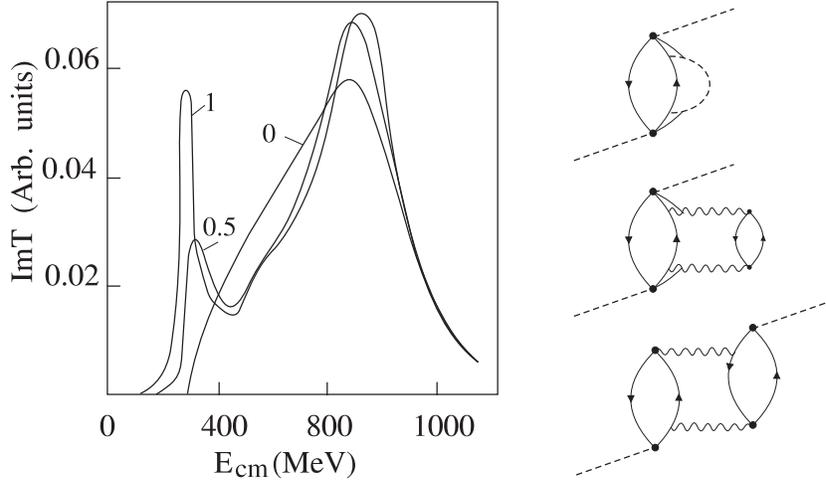}}
  \end{center}
  \caption{$\pi\pi$ strength distribution in the sigma ($I=J=0$) channel for 
  various values of $\rho / \rho_0$. The various sources of width of the
  quasi-pions are shown on the right side.}
  \label{figure5}
\end{figure}

\begin{figure}
  \begin{center}
    \mbox{\epsfxsize=11cm
          \epsfbox{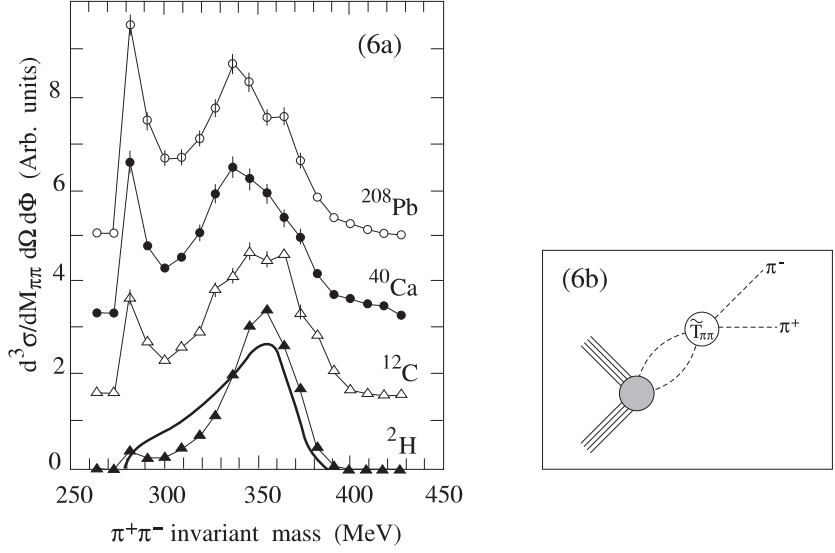}}
  \end{center}
  \caption{ (6a):Invariant mass distribution for the $\pi^+,\, \pi^+ \pi^-$
  reaction on various nuclei. The full curve is a prediction of the model of
  ref. 23 .(6b): Rescattering of the two pions through the 
  in medium $T$ matrix}
  \label{figure6}
\end{figure}

When the coupling of the pion to the $ph$ continuum is explicitely taken 
into account, a strong accumulation of stength below threshold is observed. 
In both models an instability  with respect to pion pair condensation 
is reached at a density of $1.3 \rho_0$ \cite{AOUSS}. Obviously such 
a phenomenum would have dramatic consequences for the 
nuclear equation of state. It is likely an artefact 
of the approach which 
is in conflict with some chiral symmetry constraints; let us now 
discuss this point. It is possible to relate the $I=J=0$ $\pi \pi$ 
scattering amplitude at the soft point, defined through the 
Mandelstam variables $s=m^2_\pi,\, t=0,\,u=0$, to the $\pi \pi$ sigma commutator. 
The low energy theorem, valid up to $m^4_\pi$ correction, reads
\cite{AOUSS}
\begin{equation}
T(s=m^2_\pi,\, t=0,\,u=0)={\Sigma_{\pi\pi}\over f^2_\pi}=
{1\over f^2_\pi}\,<\pi^i({\bf k}=0)|\big[Q^5, [Q^5, H]\big]| \pi^i({\bf k}=0)>  
={m^2_\pi \over f^2_\pi}
\label{LOW} 
\end{equation}
This result should be contrasted with the Weinberg result \cite{WEINB} 
at the physical threshold which gives
\begin{equation}
T(s=4 m^2_\pi,\, t=0,\,u=0)=-7\,{m^2_\pi \over f^2_\pi}
\label{PHYS} 
\end{equation}
and indicates a sign change when going off-shell. Although 
the soft-pion amplitude is not directly relevant to the $\pi \pi$ scattering
amplitude it is possible to reach kinematical conditions which are very close 
to the point where the low energy theorem (\ref{LOW}) applies. Hence, 
one may suspect that the soft-pion constraint will generate repulsion 
below threshold which is completely absent in the two above 
phenomenological models. For this reason chirally consistent  models 
have been considered such as  linear sigma model, gauged
non linear sigma model or an improved version of the 
J\"{u}lich model  \cite{AOUSS}. An other well known problem is the way 
of preserving symmetries in the unitarization procedure. 
One needs a kinematical prescription for the off-shell continuation
of the $\pi \pi$ potential in the scattering equation.  Depending 
on the prescription, the chiral properties of the type of (\ref{PHYS}) 
present at the 
tree level may or may not be conserved in the iteration. The problem 
is especially critical within the Blankenbecler- Sugar choice (the 
intermediate pion are on the energy shell but not on the mass shell). 
One possible solution is to employ a substracted dispersion relation 
to extract the real part of the $\pi \pi$ propagator. 
For instance, in the free case this is equivalent to make the replacement~:
\begin{equation}
G_{2\pi}({\bf k},E)={1\over \omega_k}\,
{1\over E^2-4\,\omega_k^2\,+i\eta}\,\,\to\,\,
{E^2 \over 4\,\omega^2_k}\,\,
{1\over \omega_k}\,
{1\over E^2-4\,\omega_k^2\,+i\eta}
\end{equation}
The presence of the explicit $E^2$ factor ensures that the contribution 
of the iteration terms are of higher order in $m^2_\pi$, thus maintaining the 
chiral properties present at the Born term level. This procedure 
is nevertheless not totally satisfactory and a very good discussion of this 
important but rather technical problem can be found in \cite{DURSO}. However,
once these chiral symmetry constraints are incorporated the 
pion pair condensation becomes strongly disfavored although some accumulation 
of strength below or near $2\pi$ threshold remains present.

Needless to say that a sizeable reshaping of the in medium $\pi \pi$ 
interaction is of utmost interest for the saturation mechanism since
correlated pion exchange is a leading piece of nucleon-nucleon attraction. 
It is thus crucial to have some experimental information about this kind 
of in medium two-pion correlation. A possible evidence for these effects is 
 provided by the $\pi-2 \pi$ data of the CHAOS collaboration 
at TRIUMF \cite{CHAOS}. As shown on fig.6a, the measured $\pi^+ \pi^-$ invariant 
mass spectrum shows a strongly $A$ dependent structure near threshold. 
This effect, absent in the $\pi^+ \pi^+$ channel which is a pure $I=2$ isospin,
indicates this is not a threshold (cusp) effect. A plausible explanation would be
the in medium rescattering (fig. 6b)  of the pions in the 
scalar-isoscalar channel, generating a sharp structure 
as explained  above. We are now in the process of incorporating  this medium 
rescattering 
\cite {BCS} on top of the detailed model of \cite{VIC} which itself (full line 
of fig. 6) cannot explain the data.

%%%%%%%%%%%%%%%%%%%%%%%%%%%%%%%%%%%%%%%%%%%%%%%%%%%%%%%%%%%%%%%%%%%
\section{The rho meson in hot and dense hadronic matter}
%%%%%%%%%%%%%%%%%%%%%%%%%%%%%%%%%%%%%%%%%%%%%%%%%%%%%%%%%%%%%%%%%%
 
The modification of vector meson masses at high temperature and/or density
is a vividly debated subject. In particular, many theoretical works have been
 devoted
to the rho meson whose mass spectrum in dense matter can be studied,
through dileptons production,
in relativistic heavy ion collisions. For instance in the
phenomenologically well established vector dominance picture (VDM), the dilepton
production rate is directly proportionnal to the imaginary part of the rho meson
propagator

\begin{equation}
{d N_{l^+ l^-}\over d^4x\, d^4q}= -{\alpha^2 \over 3 \pi^2 M^2}\,\,\,
{3 m^4_\rho\over \pi g^2}\,\,\,{1\over e^{q_0/T}-1}\,\,\,
{Im \Sigma(q_0,T)\over \big[q^2_0-m^2_\rho-\Sigma(q_0,T)\big]^2}
\label{rate}
\end{equation}
where $g$ is the VDM coupling constant and $m_\rho$ the free space rho meson
 mass.
Here in equ.(\ref{rate})
we have assumed, for simplicitely, back to back kinematics
for which the invariant mass $M$ of the pair coincides with the energy $q_0$
of the rho at rest. $T$ is the temperature of the fireball and the physics is
contained in the in medium rho self-energy $\Sigma(q_0,T)$. We will see below
how this formula can be adapted to a realistic non-equlibrium situation.
 
Using current algebra and PCAC \cite{DEY}
it is possible to show that at finite temperature
the mass of the rho meson remains unchanged to order $T^2$.
Corrections to higher order in temperature are not controlled by chiral
symmetry alone. Indeed to order $T^4$ explicit chiral models show opposite
behaviour. For instance the linear sigma model implemented with vector mesons
\cite{PIS}
predicts a decrease of the rho meson mass to order $T^4$. On the other hand
a non linear version of the sigma model \cite{SONG}  finds an increase of this
 mass to order
$T^4$. At critical temperature both agree qualitatively in that
the masses of the $\rho$ and the $a_1$ become degenerate at a value around
$1\, GeV$.
 
It is well known that in the limit of vanishing quark masses the QCD action
is scale invariant. However this approximate scale invariance is explicitely
 broken
by quantum fluctuations in the renormalization procedure. The divergence
of the associated scale current, equal to the trace of the energy-momentum
 tensor,
is given by the so called trace anomaly \cite{COL}
 
\begin{equation}
\partial_\mu D^\mu=T^\mu_{\,\mu}=m_q\, \bar q q
-{\beta(g)\over 2  g} G^a_{\mu\nu} G_a^{\mu\nu}
\end{equation}
Where $g=(\alpha_S/4\pi)^{1/2}$ is the
QCD coupling constant and $\beta(g)$ the usual Gell-Mann-Low function
governing its evolution. In addition there is a large spontaneous breaking
of scale invariance by the gluon condensate $<G.G>= <(\alpha_s/\pi)
G^a_{\mu\nu} G_a^{\mu\nu}>_{vac}\simeq (360\, MeV)^4$. It has been proposed that
this scale symmetry should be formally present in the Lagrangian of effective
theories \cite{ELL}. This can be achieved by
introducing, through simple dimensional analysis,
a  scalar dilaton field $\chi$ in the standard chiral Lagrangian, according to
 
\begin{equation}
{\cal L}={f^2_\pi\over 4}\, \left({\chi\over \chi_0}\right)^2\,
tr_f \, \partial_\mu U \partial^\mu U^\dagger\,-\,
c\,\left({\chi\over \chi_0}\right)^3 tr_f\bigg(m_q (U+U^\dagger)\bigg)
- {1\over 2} \partial^\mu \chi \partial_\mu \chi\,-\, V(\chi)+....
\end{equation}
where $U=exp\big(i\vec\tau . \vec\phi/2\big)$ and $\vec\phi$ is the pion field.
The potential $V(\chi)$ models the quantum effects responsible for the scale
anomaly \cite{SCH} and drives the vacuum expectation value
 of the $\chi$ field to a non zero value
given by the gluon condensate $\chi^4_0\sim <G.G>$.
This approach  may be questionable
since, at the principle level, a low energy effective theory should be built,
using  renormalization group techniques, from
the elimination of short wavelength quark-gluon degrees of freedom in favor of
effective hadronic degrees of freedom. In that respect it is not easy to imagine
 how
such an effective theory can keep a memory of the basic scale invariance since,
by definition,
its construction breaks scale invariance. However it leads to interesting
 scaling laws at finite temperature and/or baryonic density \cite{RHO}.
 
\begin{equation}
{\chi^*\over \chi_0}=\left({<G.G>^*\over <G.G>}\right)^{1\over 4}
=\left({<\bar q q>^*\over <\bar q q>}\right)^{1\over 3}
={f^*_\pi\over f_\pi}={m^*_V\over m_V}=....
\end{equation}
Incorporation of vector mesons, with help of the KSFR relation, 
 in such a scheme shows that their  masses
decrease with increasing $\rho$ and/or $T$ like fractional
powers of the condensates.
 
Recent dilepton production data from the CERES collaboration at CERN/SPS
\cite{CERES} show
a spectacular reshaping of the mass spectrum in $S+Au$ collisions at $200\, GeV$
per nucleon. The rho itself is depleted and there is an enhancement in the
$500 \, MeV$ invariant mass region (see fig.8). Calculations based on the
 quark-gluon
scenario turn out to be unable to reproduce the observed enhancement.
A recent transport model calculation \cite{LI}
has shown that the data can be
reproduced provided the free space rho meson mass is replaced by the in medium
dropped rho meson mass (fig.8a). In other words one is tempted to say that these
data constitute an  experimental evidence of partial chiral
symmetry restoration. Nevertheless this statement should be at least moderated
since more conventionnal many body effects \cite{CHAS}
 may explain the major part of the effect.
 
\begin{figure}
  \begin{center}
    \mbox{\epsfxsize=10cm
          \epsfbox{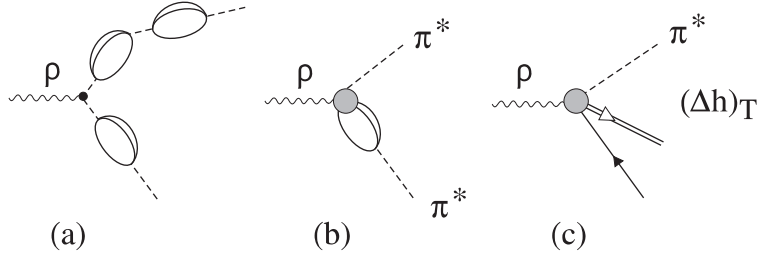}}
  \end{center}
  \caption{Medium corrections to the rho meson propagation; see explanation
in the text.}
  \label{figure7}
\end{figure}

In the invariant mass region of interest the basic process is the $\pi^+\pi^-$
 annihilation into a rho wich subsequently converts into a virtual photon.
 The first medium correction is the replacement of the bare pions by collective
 quasi pion modes $\pi^*$ ({\it i.e.} the  pionic branch)
 dressed mainly by delta-hole excitations (fig 7a). Gauge invariance in the
 medium
requires to incorporate other processes such as, among others, those 
represented on fig. 7b and 7c. The main effect of the vertex correction diagram (fig.7b) is
to kill the structure at $2 m_\pi$ coming from the dressing of pions (fig.7a).
The diagram of fig.7c represents the coupling of the rho to $\pi^* (\Delta
 h)_T$
states. The suffix  $T$ means that these Delta-hole states are the spin
transverse ones,
{\it i.e.}  those excited in photon reactions and not directly
coupled to the pionic branch.
They lead to a structure in the invariant mass region $M\ge m_\pi +\omega_\Delta
\sim 500MeV$. Similar conclusions have been reached in ref.\cite{ASA,HERR}.
 This effect accelerated by the consistent modification (of RPA type)
of the in-medium form factor
(the denominator of the propagator in equ.\ref{rate})
may provide an alternative explanation of the experimental enhancement.

\begin{figure}
  \begin{center}
    \mbox{\epsfxsize=11cm
          \epsfbox{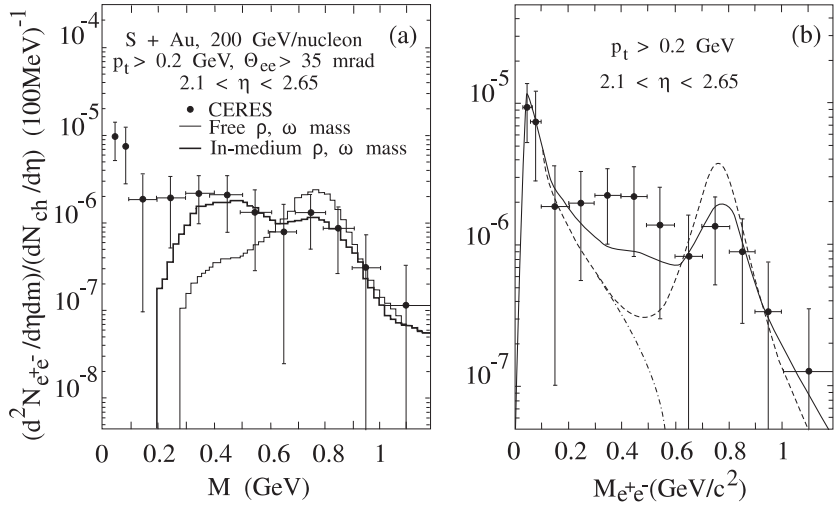}}
  \end{center}
  \caption{Effect of the in medium modified $\rho$ meson mass (8a)
 and of many body mechanisms associated with
 $\pi^*$ and  $\Delta$ (8b) compared to CERES data. The dotted curves
correspond to the bare rho calculation.}
  \label{figure8}
\end{figure}
 
We have quantitavely checked this mechanism \cite{CHAW}
by performing the full calculation
of dilepton production within the scheme of \cite{CHAS}. We have extracted the
 time evolution
of temperature and baryonic abundance from transport results \cite{LI} and
 integrated
(\ref{rate}) over time, taking into account acceptance corrections \cite{KOCH}.
For instance, at the initial time the temperature is $170\, MeV$
and the nucleon and delta densities are $1.0 \rho_0$ each. Some theoretical
improvements (inclusion of $NN^{-1},\,  N\Delta^{-1},\, \Delta\Delta^{-1}$
and thermal
effects in the corresponding bubbles) have been added to the original
calculation. The result is displayed on fig.8b where we see that
these many-body mechanisms may explain a very important part of the effect.
Nevertheless, there is still some room for those mechanisms associated
with a fundamental modification of the QCD vacuum. Thus, a consistent
 theoretical
framework, including both types of effect, has to be elaborated in view
of the next generation of dilepton experiments with the HADES detector at SIS.

\vskip 0.7 true cm
\noindent
{\bf Aknowledgement}~:  Much of the material presented in this talk
has benefited from collaboration and discussions with Z. Aouissat, J. Delorme,
M. Ericson, R. Rapp, P. Schuck and J. Wambach.

\end{document}